\documentclass[a4paper]{mem}
\usepackage{natbib}
\usepackage{graphicx}
\begin{document}
   \title{Simulations of the relativistic parsec-scale jet in 3C
   273
}

   \author{M. Perucho \inst{1},
   A.P. Lobanov \inst{2},
          \and
   J.M. Mart\'{\i} \inst{1}
}

   \offprints{M. Perucho}
\mail{manuel.perucho@uv.es}

   \institute{Departament d'Astronomia i Astrof\'{\i}sica, Universitat de Val\`encia,
                C/ Dr. Moliner 50, 46100 Burjassot (Val\`encia), Spain\\
              \and  Max Planck Institut f\"ur Radioastronomie,
              Auf dem H\"ugel 69, D-53121 Bonn, Germany
             }

\abstract{
     
  We present a hydrodynamical 3D simulation of the relativistic jet in 3C273,
  in comparison to previous linear perturbation analysis of Kelvin-Helmholtz
  instability developing in the jet. Our aim is to assess advantages and
  limitations of both analytical and numerical approaches and to identify
  spatial and temporal scales on which the linear regime of Kelvin-Helmholtz
  instability can be applied in studies of morphology and kinematics of
  parsec-scale jets.

   }
   \authorrunning{M. Perucho et al.}
   \titlerunning{Simulations of the relativistic parsec-scale jet in 3C
   273}
   \maketitle
%

\section{Introduction}

3C 273 is the brightest quasar known. Due to its relative closeness
($z=0.158$), it has become a prime target for studies in all
spectral bands aimed at understanding the AGN phenomenon \citep{courvoisier}.
 Radio emission from the jet in 3C 273 has been observed on parsec
scales using VLBI\footnote{Very Long Baseline Interferometry}
\citep{pearson,krichbaum,abraham,lobanov}. \citet[~LZ01
hereafter]{lobanov} interpreted structures observed in the
parsec-scale jet of 3C 273 as a double helix and applied linear
analysis of Kelvin-Helmholtz instability (e.g., \citealt{hardee}) to
infer the basic parameters of the flow.  \cite{asada} suggest that a
helical magnetic field could generate similar patterns in
the jet. LZ01 obtain a bulk Lorentz factor $W=2.1$, well below those
measured for superluminal components ($W=5-10$, \citealt{abraham}),
but suggest that instabilities develop in the slower underlying flow,
interpreting those fast components as shock waves inside the jet.
This result, if confirmed, would make the combination of observations
and linear stability theory a powerful tool for probing the physical
conditions in relativistic jets. We aim here to compare the results of
the analytical modelling to the structures generated in a numerical
simulation of a steady jet, with the initial conditions of the
underlying flow given in LZ01. The jet is perturbed with several
helical and elliptical modes, which correspond to the modes identified
by LZ01. Previous works by Perucho et
al. (\citeyear{peruchoa,peruchob}) have shown that numerical
simulations can be used to study the transition from the linear to the
non-linear regime, and we will apply this approach to investigate
whether the results obtained from the numerical and linear methods
can be reconciled.

\section{Numerical simulation}
 \subsection{Initial setup}

We start with a steady jet, which has a Lorentz factor $W=2.1$, density
contrast with the external medium $\eta=0.023$, sound speed
$c_{s,j}=0.53\, c$ in the jet and $c_{s,a}=0.08\, c$ in the
external medium and perfect gas equation of state (with adiabatic
exponent $\gamma=4/3$). Assuming an angle to the line of sight
$\theta = 15^\circ$ and redshift $z=0.158$ ($1\,\rm{mas}=2.43\,\rm{pc}$),
the observed jet is $169\,h^{-1}\rm{pc}$ long (with the modified Hubble constant $H_0 = 70\times h\,\rm{km}\,\rm{s}^{-1}\rm{Mpc}^{-1}$. Considering the jet
radius given in LZ01 ($0.8\, \rm{pc}$), the numerical grid is
$211\, R_j$ (axial) times $8\,R_j$ (transversal), i.e.,
$169\,\rm{pc}\times 6.4\,\rm{pc}$.

Resolution is $16\, {\rm cells}/R_j$ in the transversal direction
and $4\,{\rm cells}/R_j$ in the direction of the flow. A shear
layer of $2\, R_j$ width is included in the initial rest mass
density and axial velocity profiles to keep numerical stability of
the initial jet. Elliptical and helical modes are induced at the
inlet.
\begin{table*}
\begin{center}
\begin{tabular}{ccc|c|ccc} \label{tab:t0}
$\lambda^{obs}$& Mode & $\lambda^{theor}$&$\lambda^{sim}$ &
$\lambda^{sim}_{v_w=0.23\,c}$ & $\lambda^{sim}_{v_w=0.38\,c}$ &
$\lambda^{sim}_{v_w=0.88\,c}$\\
(mas)&&($R_j$)&($R_j$)&(mas)&(mas)&(mas)\\
\hline
2&$H_{b2}$&18.7&4&0.44&0.54&2.27\\
4&$E_{b1},H_{b1}$&37.4&25&2.7&3.37&14.3\\
12& $E_s$&21.2$^a$&50&5.5&6.7&28.5\\
\end{tabular}
\end{center}
\caption{First two columns give identified wavelengths and modes
in LZ01 ($H$ stands for helical, $E$ for elliptical, $s$ for
surface mode and $b1$ and $b2$ for first and second body modes,
respectively), third column gives the intrinsic wavelengths (see
text), in the fourth column we have written observed wavelengths
in the simulation, and the last three columns give the fourth
column wavelengths as observed depending on the wave speed. $^a$
Computed assuming it propagates with the flow speed.}
\end{table*}

Frequencies of the excited modes are computed from the observed wavelengths,
$\lambda^{obs}$, corrected for projection effects,
relativistic motion and propagation speed of the instability, $v_w$. This yields
$\omega=2\,\pi\,v_w/\lambda^{theor}$, with
\begin{equation}\label{lamb}
\lambda^{theor}=\frac{\lambda^{obs}(1-v_w/c\,\cos\theta)}{\sin\theta}\,.
\end{equation}
We use $v_w=0.23\,c$ for all body modes identified in LZ01 (see Table~1). The
surface modes are assumed to move with a speed close to that of the
jet (e.g., \citealt{hardee}). The $18\, \rm{mas}$ mode in LZ01, which
is not identified with any Kelvin-Helmholtz mode, was not included in
this simulation. Table 1 lists the modes which have been excited in
the simulation. The simulation lasted for a time $1097\, R_j/c$ (i.e.,
$\simeq 2852 \,\rm{yrs}$), and it used $\simeq 11$ Gb RAM Memory and
$8$ processors during around $30$ days in a SGI Altix 3000 computer.

\subsection{Discussion}

\begin{figure*}[!t]
   \centering
   \resizebox{\hsize}{!}{\includegraphics[bb= 205 295 400 450]{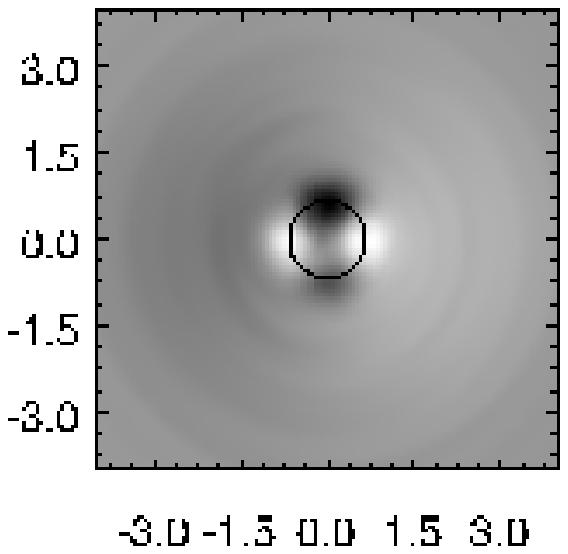}
   \includegraphics[bb= 205 295 400 450]{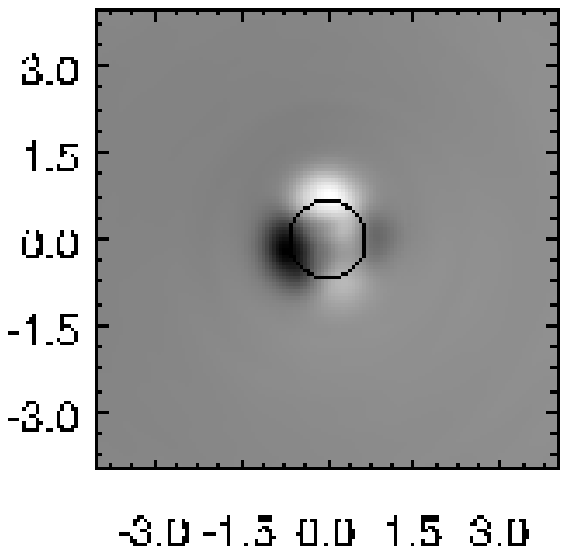}
   \includegraphics[bb= 205 295 400 450]{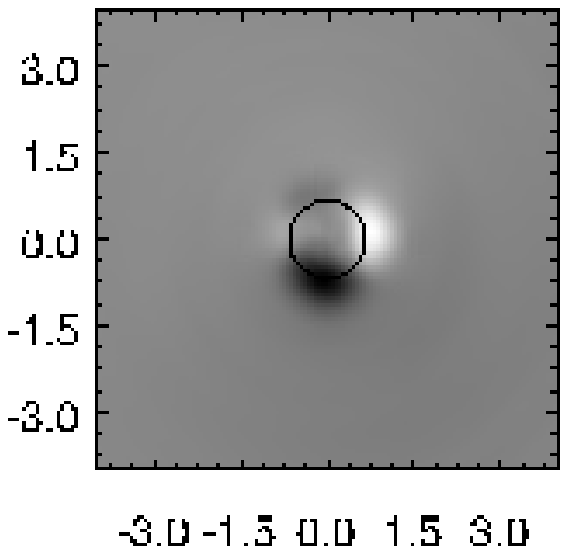}

   \includegraphics[bb= 205 295 400 450]{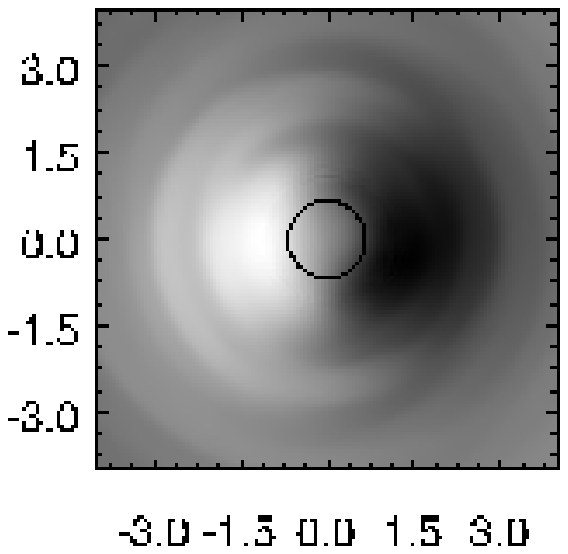}
   \includegraphics[bb= 205 295 400 450]{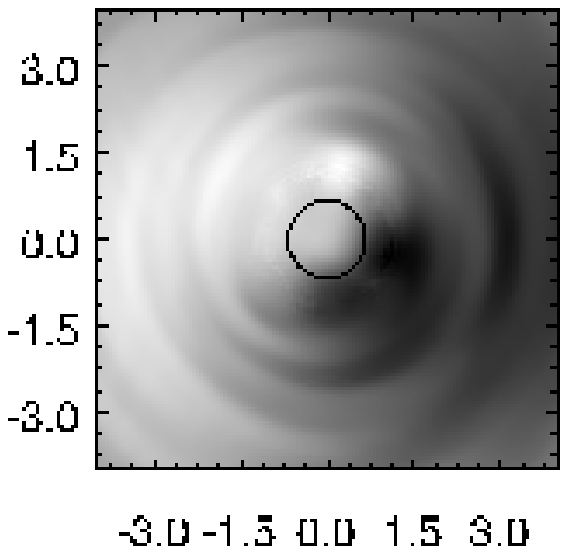}
   \includegraphics[bb= 205 295 400 450]{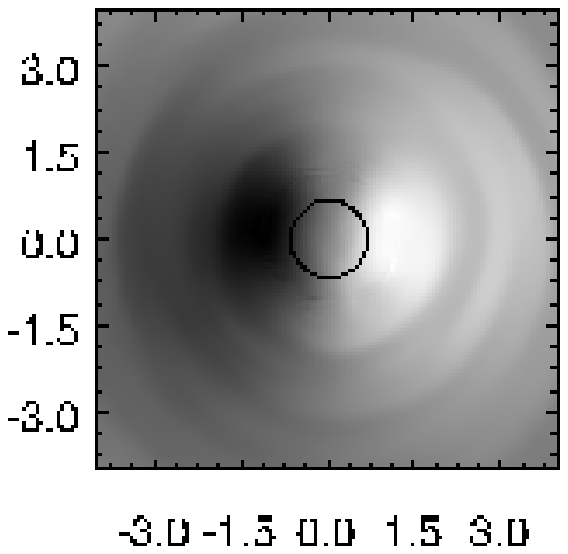}}
   \caption{Map of pressure perturbation transversal cuts in arbitrary units, values increasing from
   dark to lighter colors. Solid line indicates $v_z=0.8\,c$ contour.
   Three left panels: Cuts at $35\,R_j$, $t=70,\,140,\,200\,R_j/c$
   where elliptical mode rotation is apparent.
   Three right panels: Cuts at $105\,R_j$, $t=210,\,220,\,240\,R_j/c$
   where helical mode rotation is apparent.}
     \label{map2}
\end{figure*}

\begin{figure*}[!t]
   \centering
   \resizebox{\hsize}{!}{\includegraphics[bb= 0 0 252 40]{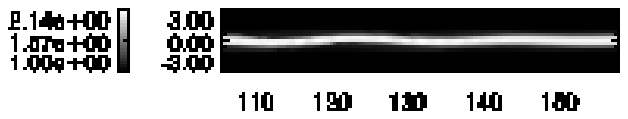}}
   \resizebox{\hsize}{!}{\includegraphics[bb= 0 0 252 40]{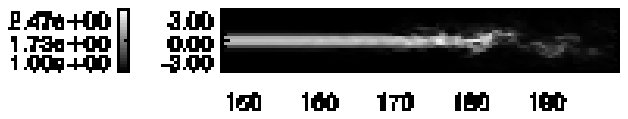}}
   \caption{Map of Lorentz factor distribution of a portion of the jet at a time before
   disruption, where a large amplitude wave is apparent (top panel, $t=320\ R_j/c$) and at
   the last frame (bottom panel, $t=1097\, R_j/c$). Coordinates are
   in jet radii. Note that the vertical scale size is increased by a factor of 4 for the sake of clarity.}
     \label{map1}
\end{figure*}

\begin{figure}
   \centering
   \resizebox{\hsize}{!}{\includegraphics{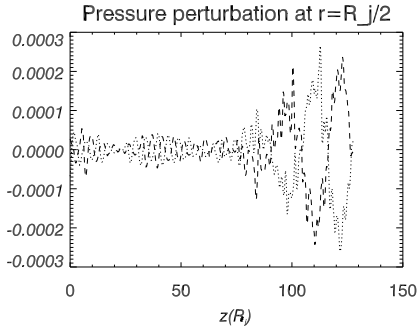}}
   \caption{Longitudinal cut of pressure perturbation at $R_j/2$ in symmetric positions
   with respect to the jet axis at $t=250\,R_j/c$. Helical (antisymmetric) structures of $4$ and $25\,R_j$
   and an elliptic (symmetric) one of $50\,R_j$ are apparent.}
     \label{pres}
\end{figure}

The simulation run can be divided in two different parts. In the
first part, the instability grows linearly. In the second part,
disruption occurs and it dominates further evolution of the flow.
In Figs. \ref{map2} and \ref{map1}, we display several transversal
cuts and axial cuts at two different times of the simulation,
respectively. The transverse cuts illustrate mode competition
showing that different modes dominate at different positions and
times in the jet. The longitudinal structures presented in Figs.
\ref{map1} and \ref{pres} indicate that the linear phase of the
instability growth is dominated by the short helical mode ($\simeq
4\, R_j$) modulated by the longer helical (antisymmetric, $\simeq
20\,R_j$) and elliptical (symmetric, $\simeq 50\,R_j$) modes.

%

The linear growth of instability ends up with the disruption of the jet caused by one
of the longest helical modes ($\simeq 20\,R_j$), at time
$t=350\,R_j/c$. After disruption, evolution of the jet is totally
influenced by this mode, and it induces perturbations that propagate slowly
backwards as a backflow in the ambient medium. The
disruption point moves outwards due to the constant injection of
momentum at the inlet and the change of conditions around the jet,
which seem to make it more stable. This point moves from $160\,
\rm{R_j}$ to $180\,\rm{R_j}$ by the end of the simulation (see Fig. \ref{map1}).

To check the consistency of the results obtained for the linear
regime of the simulation with observed structures, we need to
measure the propagation speeds of the perturbations. The time step
between subsequent frames is too large to measure these speeds
directly from the simulation data, and we use pressure
perturbations profiles for this purpose. From pressure
perturbation plots (see Fig. \ref{pres}), we estimate the velocity
of propagation of the fastest modes by measuring the position of
the perturbation front in each frame. In this way, we find
perturbations which travel with a velocity close to that of the
flow ($v_w\simeq 0.88\,c$ as an upper limit). We can also derive
the wave speed of the disruptive mode following the motion of the
large amplitude wave (see Fig. \ref{map1}) from frame to frame,
and we find $v_w\simeq 0.38\, \rm{c}$. We can associate the
former, faster perturbation, with the longer wavelength and longer
exponential growth length elliptical surface mode (see Fig.
\ref{pres}), and the latter with a shorter wavelength and shorter
exponential growth length body mode (see \citealt{hardee}). Both
measured speeds are different to that given by LZ01 ($0.23 \,c$).
This difference may be caused by an accumulation of errors in
different assumptions, or even by the excitation of a slightly
different (implying different velocity). The observer frame
wavelengths calculated for different values of the propagation
speed are given in Table~1.
%
%

%

\section{Conclusions}
We have shown that the structures found in a jet with the
physical properties of the underlying flow given in LZ01, and
perturbed with elliptical and helical modes, are of the same order
in size as those observed, if relativistic propagation effects of
the waves are taken into account. Although there is a difference
between wave speeds found in this work and that derived from
linear analysis and we do not find a unique combination of
parameters which explain observed structures, we show in Table 1
that similar wavelengths to those observed are found
for given combinations of measured wavelengths in the simulation and wave speeds.

%
%

Our simulations do not address the question of long term stability of
the jet.  The simulated jet is disrupted at distances of $\sim
170\,h^{-1}\,\rm{pc}$, which is much shorter than the observed $\sim
60\,h^{-1}\,\rm{kpc}$ extent of the jet in 3C\,273. A combination of
five different factors may be responsible for this discrepancy.
1)~Magnetic fields have not been taken into account neither in the
linear analysis, nor in the numerical simulation, but they may be
dynamically important at parsec scales.  2)~We only simulate the
underlying flow, without taking into account the superluminal
components. 3)~Initial amplitudes of perturbations cannot be
constrained accurately. 4)~Inaccuracies in the approximations used in
the linear analysis can lead to differences in physical parameters
derived. 5)~Differential rotation of the jet, shear layer thickness,
and a decreasing density of the external medium could affect the
wavelengths and speeds of the instability modes.
The combined effect of
these factors could well change the stability properties of the jet.


\begin{acknowledgements}
M.P. thanks LOC for hospitality. Calculations were performed in
SGI Altix 3000 computer \emph{CERCA} at the Servei d'Inform\`atica
de la Universitat de Val\`encia. This work was supported by the
Spanish DGES under grant AYA-2001-3490-C02. M.P. has benefited
from a predoctoral fellowship of the Universitat de Val\`encia
(\emph{V Segles}).
\end{acknowledgements}

\bibliographystyle{aa}

\end{document}